\documentstyle[11pt]{article}
\begin{document}
\renewcommand{\baselinestretch}{1.}
\baselineskip 3ex
\begin{titlepage}
\begin{flushright}
DAMTP 95-40
\end{flushright}
\vspace{3ex}
\begin{center}
{\LARGE \bf
\centerline{ $O(N)$ and $RP^{N-1}$ Models }
\centerline{ in Two Dimensions }
}
\vspace{0.65 cm}
{\large     Martin Hasenbusch }

\vspace{0.17 cm}
{\it  DAMTP, Silver Street, Cambridge, CB3 9EW, England}
\end{center}
\setcounter{page}{0}
\thispagestyle{empty}
\begin{abstract}\normalsize
I provide evidence that the 2D $RP^{N-1}$ model for $N \ge 3$ is equivalent to 
the $O(N)$-invariant non-linear $\sigma$-model in the continuum limit.  
To this end, I mainly study  particular versions of the models, to be called 
constraint models.
I prove that the constraint $RP^{N-1}$ and $O(N)$ models
are equivalent for sufficiently weak coupling.
Numerical results for the step-scaling function of the running coupling 
$\bar{g}^2= m(L) L$ are presented.
The data confirm that the
constraint $O(N)$ model is in the same universality
class as the $O(N)$ model with standard action. 
I show that the differences in the finite size scaling curves of 
$RP^{N-1}$ and $O(N)$ models 
observed by Caracciolo et al. can be explained as a
boundary effect. 
It is concluded, in contrast to Caracciolo et al.,
 that  $RP^{N-1}$ and $O(N)$ models share a unique universality class.
\end{abstract}
\nopagebreak
\begin{flushleft}
\vspace{0.5 cm}
DAMTP 95-40
July 1995
\end{flushleft}
\vspace{1ex}
\end{titlepage}

\newpage
%%%%%%%%%%%%%%%%%%%%%%%%%%%%%%%%%%%%%%%%%%%%%%%%%%%%%%%%%%%%%%%%%%%%%%%%%%%%%%%
%                          BODY OF THE PAPER                                  %
%%%%%%%%%%%%%%%%%%%%%%%%%%%%%%%%%%%%%%%%%%%%%%%%%%%%%%%%%%%%%%%%%%%%%%%%%%%%%%%
\section{Introduction}
Motivated  by the close analogy with non-Abelian lattice gauge theories in four dimensions
non-linear $\sigma$-models in two dimensions have been studied intensively during
the last 20 years. 
Most important both types of models were found to be 
asymptotically free \cite{polyakov}.

Starting from the early eighties so called $RP^{N-1}$ models were discussed.
The spins of these models are elements of the real projective space in $N$ 
dimensions. This space can be thought of as a sphere $S^{N-1}$ where opposite 
points are identified. Hence in perturbation theory
the $RP^{N-1}$ model is equivalent with the $O(N)$-invariant $\sigma$-model.
The fact however that the real projective space is not simply connected 
gives rise to topological defect structures similar to vortices in the 
2D $XY$ model. 
The questions discussed in the literature are whether these defects  
induce a phase transition at a finite coupling or whether
these non-perturbative effects survive in the weak coupling limit.  

The lattice action of the $RP^{N-1}$ model mostly discussed 
is 
\begin{equation}
S=- \beta \sum_{<xy>} (\vec{s}_x \vec{s}_y)^2 \;\;,
\end{equation}
where $<xy>$ is a pair of nearest neighbour points on the lattice and
$\vec{s}$ is a unit vector in $R^N$. 
An alternative way to identify $-\vec{s}_x$ and $\vec{s}_x$ 
is to introduce an $Z_2$ gauge field
\begin{equation}
S=- \beta \sum_{<xy>} z_{<xy>} \vec{s}_x \vec{s}_y  \;\;,
\end{equation}
where $z$ takes the values $1$ or $-1$. 

Similar models have been introduced to 
describe orientational phase transitions in
nematic liquid crystals \cite{Lebwohl}. These models have mainly been 
studied in three dimensions, where a weak first order phase transition is 
found (see ref.~\cite{green} and references given in \cite{chic}).

The numerical study of  $RP^{N-1}$ models in 2D gave rise to much controversy. 
Some authors \cite{sorin,torino,Kunz} find that their results are
consistent with a phase transition at a finite coupling, while others
doubt the existence of a phase transition but still see strong crossover
effects between the strong and the
 weak coupling regime \cite{green,sorin2,chic}.

Recently Caracciolo et al. \cite{sokal1,sokal2,sokal3} argued that there is 
no phase transition in the $RP^{N-1}$ models. They
claimed, based on a finite size scaling analysis, that the $RP^{N-1}$ models
however have a weak coupling limit distinct
 from that of the $O(N)$-invariant $\sigma$-models.
They claim even further, that a whole sequence of universality classes 
can be obtained from mixed models. 

In the following I will give evidence that rules out the scenario presented
in \cite{sokal1,sokal2,sokal3}. For a particular type of the 
action of the $O(N)$ and the $RP^{N-1}$ model I will show that the models are
exactly equivalent for sufficiently small coupling.  
I discuss the scaling properties of vortices of the $RP^{N-1}$ model with
 standard action in the weak coupling regime.
 I give numerical results for the step scaling function introduced in 
ref.~\cite{luscher} that 
supports that the constraint model gives the same universal results as 
the standard action. 
I show that the differences in the finite size scaling curves for the $RP^{2}$
and the $O(3)$ model found 
in refs.~\cite{sokal2,sokal3} can be explained as a boundary effect. 

\section{The Constraint $O(N)$ and $RP^{N-1}$ Models}

Let me first define the models. The field variable $\vec{s}_x$ 
is in both cases a unit
vector in $R^N$. In the case of  the $O(N)$-invariant model the Boltzmann
weight of a configuration is equal to $1$ if 
\begin{equation} 
 \vec{s}_x  \vec{s}_y > C
\end{equation} 
for all nearest neighbour pairs of sites $<xy>$. Else the Boltzmann weight is 
equal to $0$. 

In the case of the $RP^{N-1}$ model $-\vec{s}_x$ and
 $\vec{s}_x$ are identified and the 
constraint on the field configuration is given by
\begin{equation} 
 |\vec{s}_x  \vec{s}_y| > C
\end{equation} 
for all nearest neighbours  $<xy>$. 
Equivalently one might introduce a gauge field $z_{<xy>}$ taking the values 
$-1$ or $1$. 
\begin{equation} 
 z_{<xy>} \vec{s}_x  \vec{s}_y > C \;\; .
\end{equation} 

In the following 
I shall show that the constraint $O(N)$-invariant model and the constraint
$RP^{N-1}$ model
 are equivalent for $C > \cos(\pi/4)$.
Let us consider a lattice where all closed paths are contractible,
i.e. all closed paths
 can be shrinked to an elementary plaquette by removing single
plaquettes sequentially.  
A hyper-cubical square lattice with open boundary conditions is an example for 
such a lattice. 
 
 Consider the class of $2^V$, where $V$ is the number of lattice points,
 configurations that arise from a given 
 configuration $\vec{s}_x$ by taking
 either
 $+\vec{s}_x$ or $-\vec{s}_x$  at each lattice point. 
 Since
\begin{equation} 
 |\vec{s}_x \vec{s}_y| = |(-\vec{s}_x) \vec{s}_y| = |\vec{s}_x  (-\vec{s}_y)|
                 =|(-\vec{s}_x)  (-\vec{s}_y)| 
\end{equation} 
 all configurations in such a class are either allowed or 
 forbidden $RP^{N-1}$ configurations.  Obviously a class of configurations
 that is forbidden under the  $RP^{N-1}$ constraint contains no configuration 
 that is allowed under the $O(N)$ constraint (with the same $C$).
  In the following I will
 demonstrate that for $C > \cos(\pi/4) $ a class of configurations that is 
 allowed under the $RP^{N-1}$ constraint contains exactly $2$ configurations
 allowed under the $O(N)$ constraint, and therefore the partition functions
 are equal up to a trivial factor $2^{V-1}$.   

 Take one configuration out of an allowed class of $RP^{N-1}$ configurations.
 Pick one site $x$. 
 Replace the spins on the other sites by
\begin{equation} 
\vec{s}_y \, {}' =\vec{s}_y  
\prod_{<uv> \in \mbox{ path}(x,y)}   \mbox{sign}(\vec{s}_u \vec{s}_v)  \;\; . 
\end{equation} 
The result of this construction is independent of the paths chosen  if 
\begin{equation} 
\prod_{<uv> \in \mbox{closed path}}   \mbox{sign}(\vec{s}_u \vec{s}_v) = 1 
\end{equation} 
for all closed paths on the lattice.  For elementary loops consisting of 
four lattice-points this is the case for $C > \cos(\pi/4) $. 
All other paths can be successively built up out of elementary
loops, since we have chosen a simply connected lattice topology.
 When adding an elementary loop the sign of a loop is conserved
             since the sign of the 
             product of the new links in the path is the same as for the 
             old links. Hence the sign  of any closed path is $1$ for 
             $C > \cos(\pi/4)$.

The idea behind this proof has been discussed for an action similar
to that in eq.~(2) by Caselle and Gliozzi  \cite{torino}. However for that
action the rigorous proof for a pure gauge in the weak coupling limit
is missing.   

In Monte Carlo simulations one typically uses periodic boundary conditions, 
which leads to the lattice topology of a torus. Here loops exist that wind 
around the torus and hence cannot be contracted to an elementary loop. 

In order to avoid configurations that are allowed under the $RP^{N-1}$
 constraint but not for $O(N)$
  one has to require $C > \cos(\pi/L) $ where $L$ is the 
extension of the lattice in units of the lattice spacing. It is important 
 to note that such boundary effects might well survive the continuum limit  
in a finite size scaling analysis.   
 However this boundary effect can be reproduced by proper boundary 
 conditions imposed upon the $O(N)$ model. 
For $C >  \cos(\pi/4)$ a constraint
$RP^{N-1}$  model on a periodic lattice is equivalent 
to a constraint  $O(N)$ model
 with fluctuating boundary conditions.  Fluctuating
boundary conditions mean that
in the partition function one sums over periodic as well as anti-periodic
boundary conditions. In the case of anti-periodic boundary conditions one
identifies $\vec{s}(0,y) = - \vec{s}(L,y)$, $\vec{s}(L+1,y) = - \vec{s}(1,y)$
 $\vec{s}(x,0) = - \vec{s}(x,L)$ and $\vec{s}(x,L+1) = - \vec{s}(x,1)$. 

\section{Scaling of the Vortex Density for the Standard Actions}
 For the standard actions of the $RP^{N-1}$ model similar arguments 
 apply. In the limit $\beta \rightarrow \infty$ the energy of a vortex 
 should win against the entropy and vortices should play no role
 in the continuum limit of the theory. 

 Let us identify a frustrated plaquette in eq.~(2) with the center of a 
 vortex. 
 The classical solution of the $\vec{s}$-field for a fixed gauge field
 with two frustrated plaquettes has an energy proportional to 
 $\ln r$ where $r$ is the distance in between these two frustrated
 plaquettes.
  Hence one can find a finite $r_0$ such that the energy is larger than
  $2 b_0 + \epsilon$,  where  $b_0$ is the leading coefficient in the
  perturbative $\beta$-function. 
 Therefore the density of vortex pairs with a distance larger than
  $r_0$ dies out faster than the square of 
 the inverse correlation length. Hence they can not play a role in the 
 continuum limit of the theory. 

\section{Numerical Results for the Constraint Models}
    In this section I show that the constraint $O(N)$ model
   reproduces universal results of the $O(N)$-invariant $\sigma$-model. 
   Therefore I compute the step scaling function of ref. \cite{luscher}
   for three different values of the running
   coupling and compare the result with that of
   ref.~\cite{luscher} obtained with the standard action.
   Furthermore I estimate the correlation length at $C=\cos(\pi/4)$  using the
   running coupling and also  measure the correlation length
   for both the $O(N)$ and the $RP^{N-1}$ model for $C < \cos(\pi/4)$
    to check the importance of
   defects  in the generation of the mass in the $RP^{N-1}$ case.

 The running coupling of ref.~\cite{luscher} is defined by
 \begin{equation}
 \bar{g}^2 = \frac{2}{N-1} m(L) L \; , 
 \end{equation}
 where $m(L)$ is the mass gap on a lattice with extension $L$ in spatial
 direction. 
 The $\beta$-function for the running coupling $\bar{g}^2$
 is  given by \cite{luscher}
\begin{equation}
 \beta(g^2) = - \frac{N-2}{2 \pi} \bar{g}^4 - \frac{N-2}{(2 \pi)^2} \bar{g}^6 
               -\frac{(N-1)(N-2)}{(2 \pi)^3} \bar{g}^8  ... \;\; .
\end{equation}
   The step scaling function $\sigma(s,u)$ is the discrete version of the
   $\beta$-function. It gives the value of the coupling after  
   change of $L$ by a factor of $s$ starting from a coupling $u$.

   The simulation was done using the evident modification of the single
   cluster algorithm \cite{ullisingle}.  A bond $<xy>$ is called deleted if
   after the reflection of one of the spins $\vec{s}_x$ or $\vec{s}_y$
   the constraint $\vec{s}_x \vec{s}_y <  C$ is still satisfied.

 A proof of ergodicity is given in the appendix.
   The simulation results listed in table 1 are based on about $10^7$ single
   cluster updates.
 The correlation function was measured using the
   cluster-improved
   estimator \cite{ullio3} . The mass was extracted from the 
   correlation function at distance $L$ and $2L$. 

   Fitting the data of table 1 to an Ansatz
   \begin{equation}
   \Sigma(2,u,a/L) =  \sigma(2,u) + c \; (a/L)^2 \;\; .
   \end{equation} 
   I obtain $\sigma(2,1.0595)=1.2589(10)$ from $L/a \ge 16$,
   $\sigma(2,0.8166)=0.9150(8)$ 
   from $L/a \ge 8 $ and $\sigma(2,0.7383)= 0.8159(8)$ from $L/a \ge 8$. 
   These results can be compared with the step scaling
   function  obtained in ref.~\cite{luscher}  $\sigma(2,1.0595)=1.2641(20)$,
   $\sigma(2,0.8166)=0.9176(8)$ and $\sigma(2,0.7383)=0.8166(9)$. The slight
   disagreement  (about 2 standard deviations) might well 
   be explained by deviations
   of the corrections to finite size scaling from the fit-Ansatz chosen.

The exact prediction for the mass gap given by
\cite{hasenfratz}
\begin{equation}
\frac{m}{\Lambda_{\overline{MS}}} = \frac{8} {\mbox{e}}
\end{equation}
for $N=3$ and
the conversion factor for the $\Lambda$ parameters 
\begin{equation}
 \Lambda = \frac{e^{-\Gamma'(1)}}{4\pi} \Lambda_{\overline{MS}}
\end{equation}
given in ref.~\cite{luscher} allows us to give an estimate for the infinite
volume correlation length based on the measurement of the correlation length
 on a  finite lattice.
   Taking the Monte Carlo result for the running coupling given in table 1
   I obtain $\xi = 0.7 \times 10^5 $ as 
   estimate for the correlation length at $C=0.55$ and $\xi = 0.6 \times 10^9 $
   as estimate at $C=\cos(\pi/4)$,
 where the $O(3)$ and $RP^{2}$ constraint 
   models become identical. 

   In addition I performed some simulations for both the constraint 
   $O(N)$ model and the constraint $RP^{N-1}$ model
   at smaller $C$ values such that
   the correlation length $\xi$ is much smaller than the lattice size $L$. 
   Here I adopted the definitions used in refs.~\cite{sokal1,sokal2,sokal3}. 

   The correlation function in the vector-channel is defined by
   \begin{equation}
   G_v(x,y) = \langle \vec{s}_x \vec{s}_y \rangle  \;\; .
   \end{equation}
   Since naively this quantity vanishes identically under the symmetries of the 
   $RP^{N-1} $ model one considers the tensor channel with the correlation
   function
    \begin{equation}
   G_t(x,y) = \langle (\vec{s}_x \vec{s}_y)^2 \rangle -  \frac{1}{N} \;\; .
   \end{equation}
   Starting from these definitions of the correlation function one obtains 
   the susceptibility 
   \begin{equation}
     \chi = \frac{1}{V} \sum_{x,y} G(x,y)
   \end{equation}
   and 
   \begin{equation}
     F= \frac{1}{V} \sum_{x,y} \cos(\frac{2 \pi}{L} k (x-y))) \;\; G(x,y)
   \end{equation}
   with $k=(1,0)$ or $k=(0,1)$. 

   The second moment correlation length is now defined as 
   \begin{equation}
   \xi =  \frac{\sqrt{\chi/F-1}}{2 \sin(\pi/L)} \;\; .
   \end{equation}

   In table 2 some results for the constraint $O(3)$  model are given. 
   It is remarkable that  already for $C=0$ the correlation length is 
   larger than ten. The ratio of the correlation length 
    in the vector and the tensor channel is about $\xi_v/\xi_t = 3.3(1)$. 

In table 3 my results for the constraint $RP^2$ model are summarized. 
 At $C=0.55$ there is a factor of about 
   $10^3$ in between the correlation lengths of the $O(3)$ and the $RP^2$
  model. 
   This means that it is practically impossible to see the true asymptotic 
   behaviour of the constraint $RP^{2}$ model in  a computer simulation. 

\section{Finite Size Scaling and Universality}
 In this section I shall demonstrate that the difference in the finite 
 size scaling curves observed in \cite{sokal1} and \cite{sokal2} can be
 explained in part by the boundary effect discussed above. 
 I simulated the $O(3)$-invariant model with the standard action
 on a square lattice using fluctuating
 boundary conditions in both lattice-directions. 
 For the updates of the boundary
 conditions  I used the boundary flip algorithm proposed in ref.~\cite{fluct}
 for the Ising model 
 and generalized to $O(N)$ vector models in ref.~\cite{fluctxy}. 

 I performed runs at $\beta=1.4$,  $\beta=1.5$  and
 $\beta=1.6$. The true correlation
 lengths for these $\beta$ values are $\xi = 6.90(1)$, $\xi = 11.09(2)$ and
 $\xi = 19.07(6)$, respectively \cite{ullio3}.
 I used lattice sizes ranging from $L=6$ to $L=128$. Throughout I performed 
 100000 measurements. I performed a measurement after
 one boundary-flip update for each 
 direction and roughly $cluster$-$size/lattice$-$size$ 
 standard single cluster updates.
% In figure 1 the dimensionless 
% quantity $\chi_t(2L)/\chi_t(L)$ is plotted as a function of $\xi/L$. 
% In order to compare with fig. 2a  of ref.~\cite{sokal3} one has to take 
% the factor $\xi_v/\xi_t = 3.3(1) $ into account.  
% The fluctuating boundary conditions remove the characteristic dip visible
% in the finite size scaling curve of fig. 2b of ref.~\cite{sokal3}
% ($O(3)$ model with periodic boundary conditions). The 
% curve qualitatively looks much like fig. 2a of ref.~\cite{sokal3}
% ($RP^2$ like models).  
% However the slope of the curve in fig. 2a of ref.~\cite{sokal3}
% is much steeper
% for small $\xi/L$ than that of fig.~1. This shows that
% the results given in fig.~2a of ref.~\cite{sokal3} are strongly effected 
% by corrections to scaling.
In figure 1 the dimensionless
 quantity $\chi_t(2L)/\chi_t(L)$ is plotted as a function of $\xi/L$.
 I give the results for fluctuating boundary conditions (circles) and
 for comparison the results with periodic boundary conditions (diamonds).
 In order to compare with fig. 2  of ref. \cite{sokal3} one has to take
 the factor $\xi_v/\xi_t = 3.3(1) $ into account.
 My result for periodic boundary conditions is consistent with that given
 in fig. 2b of ref. \cite{sokal3}.
 The fluctuating boundary conditions remove the characteristic dip visible
 in the finite size scaling curve for periodic boundary conditions.
 The finite size scaling curve for periodic boundary conditions
 looks  qualitatively much like that of fig. 2a of ref. \cite{sokal3}
 ($RP^2$ like models).
 However the slope of the curve in fig. 2a of ref. \cite{sokal3}
 is much steeper for large $\xi/L$ than that of fig. 1. This shows that
 the results given in fig. 2a of ref. \cite{sokal3} are effected
 by strong corrections to scaling due to vortices.

\section{Is There a Phase Transition?} 
In order to understand the phase-structure of the $RP^{N-1}$ models one might,
in analogy with the $KT$ scenario of the $XY$ model \cite{KT}, 
discuss the RG-flow of the models in a 2 dimensional parameter space. 
In addition to the coupling $g^2$ one might 
introduce a coupling parameter $\mu$
for the plaquette-term, controlling the density of vortices.
\begin{equation}
 S = - \frac{1}{g^2} \sum_{<xy>} z_{xy} \vec{s}_x \vec{s}_y + \ln \mu 
 \sum_p z_p \;\; , 
\end{equation} 
where $z_p = \prod_{<xy> \in p} z_{xy}$. 

I will make no attempt here to derive the RG flow-equations. However 
certain qualitative features and their consequences seem to be evident:

a) For $(g^2,0)$ the standard $\beta$-function of the $O(N)$ model
 is recovered. 

b) Vortices cause disorder. Therefore a non-vanishing $\mu$ should  amount to 
a positive contribution in the derivative of $g^2$
 with respect to the logarithm
of the cutoff scale and hence accelerate the flow towards strong coupling.

Statement a) rules out  that a possible phase transition in $RP^{N-1}$ is
$KT$ like, since the fixed-point of the $KT$-transition is purely Gaussian.  
Furthermore statement b) rules out any fixed-point that might occur
at a finite $\mu$.

Still we have to explain why Monte Carlo simulations and strong coupling 
expansions seem to be in favour of a phase transition. It seems plausible
that in analogy with the $KT$-flow equations $\mu$ is irrelevant for small
coupling $g^2$ but becomes relevant above some threshold value $g^2_t$.  
That means above $g^2_t$ the RG-trajectories are driven off from the 
renormalized trajectory of the $O(N)$-invariant model.

\section{Conclusions}
I have proven that the constraint $O(N)$ and constraint $RP^{N-1}$ model become 
equivalent for $C > \cos(\pi/4) $.  Using the renormalized coupling  
$\bar{g}^2 = m(L) L$ I estimated the correlation length at $C=\cos(\pi/4) $
to be about $\xi = 0.6 \cdot 10^9$
 for both models with $N=3$. For $C$-values 
being smaller, such that $\xi << 1000$,  the models display huge differences.   
 This means that the asymptotic behaviour of the constraint
$RP^{2}$ practically can 
 not be observed in a computer simulation.  I argue that  
 a similar scenario holds for models with a standard action.
As $\beta \rightarrow \infty$  vortices in the $RP^{N-1}$ model 
vanish and the $RP^{N-1}$ becomes equivalent to an $O(N)$ model by the 
virtue of a gauge fixing. 

On lattices with periodic boundary conditions
one has to notice that paths winding 
around the lattice are not contractible. The effect of such loops in the 
$RP^{N-1}$ model amount to fluctuating boundary conditions in the 
equivalent $O(N)$ model.
I demonstrated numerically that this fact partially
 explains the differences
found in the finite size scaling curves  for the $O(3)$ and $RP^{2}$  models
observed in refs.~\cite{sokal2,sokal3}.

\section{Note added}
 My conclusions are confirmed by the work of 
 F. Niedermayer, P. Weisz and D.-S.  Shin \cite{heplat},
 which I found today on the 
 hep-lat bulletin-board.

\section{Appendix}
     In the following I prove that the single cluster algorithm applied
     to the constraint $O(N)$-invariant model is ergodic.  

     It is sufficient to show that any allowed configuration can be transformed
     in a finite number of  cluster-update steps to the configuration
     $\vec{s}=(1,0,...,0)$ for all sites.

     Let us consider a $N=2$ ($XY$)
 model with a bond dependent constraint $C_{<xy>}$.
     Assume that the spins are distributed in an angle range $[0,\alpha_k]$
     with respect to the 1-axis.
     (The largest range to start with is $[0,2\pi]$.)

     Take a reflection axis which
     has an angle $\alpha_k/2$ with the $1$-axis.
     Per construction none of the sites $x$
     with $\phi_x > \alpha_k/2$ is connected
     via a frozen bond with a site $y$ with $\phi_y < \alpha_k/2$.
     Hence  all spins can be moved into the range $[0,\alpha_k+1]$
     with $\alpha_{k+1}=\alpha_{k}/2$ using  a finite number (smaller or equal
     the number of sites) of cluster updates.
     Iterating this process, in a finite number of steps
     all spins can be put into the range
     $[0,\mbox{min} \arccos(C_{<xy>})]$.
     Now take for each site a reflection axis with $\alpha_x = \phi_x/2$. 
     Per construction all these clusters are single site clusters. 

     We hence constructed a sequence of a finite  number of cluster updates
     that transforms an arbitrary configuration to the $s=(1,0)$ for all sites
     configuration.

     For general $N$ this procedure can be iterated.
     Consider the $N^{th}$ and $(N-1)^{th}$
     component as an embedded XY model. Remove the $N^{th}$ component.
     Go ahead untill  $s=(1,0,...,0)$ for all sites is reached.

\section*{Acknowledgements}

I would like to thank M. Caselle, I.T. Drummond, R.R. Horgan, and K. Pinn
for discussions. This work was supported by the Leverhulme Trust   
under grant  16634-AOZ-R8 and by PPARC. 

%%%%%%%%%%%%%%%%%%%%%%%%%%%%%%%%%%%%%%%%%%%%%%%%%%%%%%%%%%%%%%%%%%%%%%%%%%%%%%%
%                           REFERENCES                                        %
%%%%%%%%%%%%%%%%%%%%%%%%%%%%%%%%%%%%%%%%%%%%%%%%%%%%%%%%%%%%%%%%%%%%%%%%%%%%%%%

\vspace{4cm}

{\large \bf Figure captions}

\vspace{2cm}
Figure 1:
 The dimensionless quantity $\chi_t(2L)/\chi_t(L)$ is given as a function
 of $\xi_v/L$ for the $O(3)$ model. The data points with the circles
 are obtained with fluctuation boundary conditions  while those with diamonds
 are obtained with periodic boundary conditions.

\newpage
%%%%%%%%%%%%%%%%%%%%%%%%%%%%%%%%%%%%%%%%%%%%%%%%%%%%%%%%%%%%%%%%%%%%%%%%%%%%%%%
%                             TABLES                                          %
%%%%%%%%%%%%%%%%%%%%%%%%%%%%%%%%%%%%%%%%%%%%%%%%%%%%%%%%%%%%%%%%%%%%%%%%%%%%%%%

\begin{table}
\caption{ The renormalized coupling $\bar{g}^2 $ from the constraint 
O(3) model. $C$ gives the value of the constraint. $L/a$ and $L'/a$ 
are the lattice extensions in spacial direction.
        }
         \label{tab1}
\begin{center}
\begin{tabular}{|crrcc|}
\hline
 $C$  & $L/a$  &  $L'/a$  &  $ \bar{g}^2(L)$   & $\bar{g}^2(L')$ \\
\hline
 0.0515 & 4    &    8     &   1.0595(2)        & 1.2623(3)  \\
 0.0820 & 5    &   10     &   1.0595(2)        & 1.2564(3)  \\
 0.1047 & 6    &   12     &   1.0595(2)        & 1.2542(3)  \\
 0.1225 & 7    &   14     &   1.0595(2)        & 1.2540(2)  \\
 0.1371 & 8    &   16     &   1.0595(2)        & 1.2542(4)  \\
 0.1607 & 10   &   20     &   1.0595(2)        & 1.2545(3)  \\
 0.1786 & 12   &   24     &   1.0595(2)        & 1.2542(4)  \\
 0.2058 & 16   &   32     &   1.0595(2)        & 1.2559(3)  \\
 0.2255 & 20   &   40     &   1.0595(2)        & 1.2569(3)  \\
 0.2637 & 32   &   64     &   1.0595(2)        & 1.2582(6)  \\
\hline
 0.1992 & 4    &    8     &   0.8166(2)        & 0.9358(3)  \\
 0.2413 & 6    &   12     &   0.8166(2)        & 0.9234(2)  \\
 0.2663 & 8    &   16     &   0.8166(2)        & 0.9186(3)  \\
 0.2835 & 10   &   20     &   0.8166(2)        & 0.9171(2)  \\
 0.2967 & 12   &   24     &   0.8166(2)        & 0.9167(3)  \\
\hline
 0.2538 & 4    &    8     &   0.7383(2)        & 0.8373(2)  \\
 0.2924 & 6    &   12     &   0.7383(2)        & 0.8246(2)  \\
 0.3147 & 8    &   16     &   0.7383(2)        & 0.8197(3)  \\
 0.3299 &10    &   20     &   0.7383(2)        & 0.8191(4)  \\
 0.3416 &12    &   24     &   0.7383(2)        & 0.8174(3)  \\
\hline  
 0.55       &16&   32     &  0.4491(2)         &   0.4759(4) \\
$1/\sqrt{2}$&16&   32     &  0.2654(1)         &   0.2746(1)\\
\hline
\end{tabular}
\end{center}
\end{table}

\begin{table}
\caption{ The second moment
 correlation length in the vector ($\xi_v$) and tensor
 ($\xi_t$) channel for
 the constraint $O(3)$-invariant 
vector model for various values of the constraint $C$.  
        }
         \label{tab2}
\begin{center}
\begin{tabular}{|crcc|}
\hline
$C$ & L & $\xi_v$& $\xi_t$  \\
\hline
0.00  & 64 & 11.20(5)&3.29(6)\\
0.10  &128 & 23.3(2) & 6.9(2)  \\
0.2255&400 & 76.7(4) & 24.0(6) \\
%0.00  & 64 & 11.20(5)& 168.3(6) & 3.29(6)& 1.646(5) \\
%0.10  &128 & 23.3(2) & 597.(5.) & 6.9(2) & 4.19(4) \\
%0.2255&400 & 76.7(4) &4968.(25.)& 24.0(6)&21.95(12) \\
\hline
\end{tabular}
\end{center}
\end{table}

\begin{table}
\caption{The second moment correlation length in the tensor  channel $\xi_t$
  for the constraint $RP^{2}$ model for various values of the constraint $C$.
        }
         \label{tab3}
\begin{center}
\begin{tabular}{|crc|}
\hline
$C$ & L & $\xi_t$ \\
\hline
0.50 &64 & 4.72(2)  \\
0.51 &64 & 5.66(2)  \\
0.52 &128& 7.10(6)  \\
0.53 &128& 9.06(5)  \\
0.55 &128&16.52(7)  \\
%0.50 &64 & 4.72(2) & 11.91(2) \\
%0.51 &64 & 5.66(2) & 16.03(3) \\
%0.52 &128& 7.10(6) & 22.64(4) \\
%0.53 &128& 9.06(5) & 33.60(7) \\
%0.55 &128&16.52(7) & 89.5(3)  \\
\hline
\end{tabular}
\end{center}
\end{table}

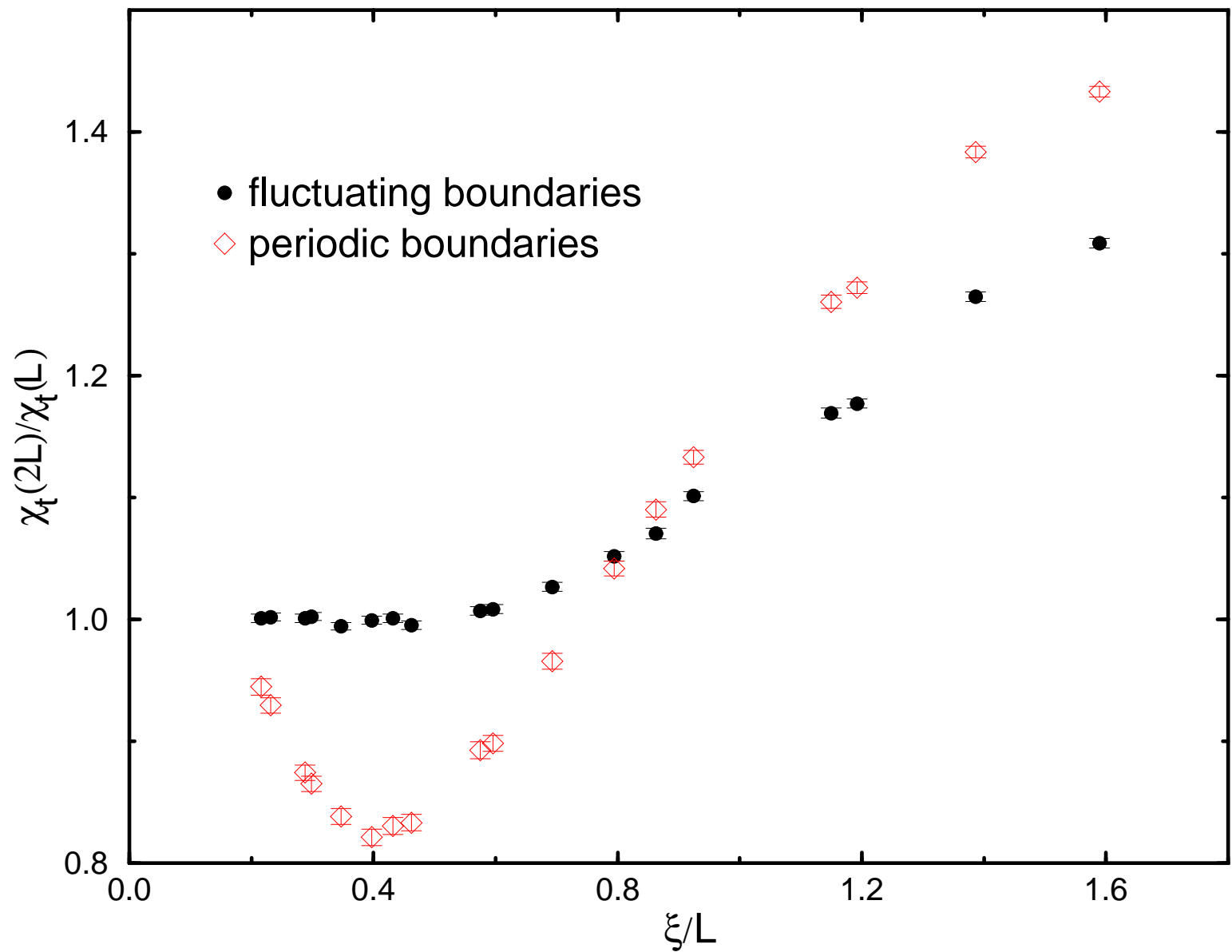
\end{document}